\documentstyle [12pt]{article}

\newcommand{\beq}{\begin{equation}}
\newcommand{\dd}{\partial}
\newcommand{\eeq}{\end{equation}}
\newcommand{\bea}{\begin{eqnarray}}
\newcommand{\eea}{\end{eqnarray}}
\newcommand{\fnu}{f^{(\nu)}}
\newcommand{\fn}{f^{(n)}}

\begin{document}

\baselineskip 7.5 mm

\def\thefootnote{\fnsymbol{footnote}}

\begin{flushright}
\begin{tabular}{l}
CERN-TH/98-192 \\
astro-ph/9806205
\end{tabular}
\end{flushright}

\vspace{12mm}

\begin{center}

{\Large \bf
Neutrino transport: no asymmetry in equilibrium 
}
\vspace{18mm}

\setcounter{footnote}{0}

{\large
Alexander Kusenko,}$^1$\footnote{ email address: Alexander.Kusenko@cern.ch}
{\large
Gino Segr\`{e},$^2$\footnote{email address: segre@dept.physics.upenn.edu} }
{\large 
and Alexander Vilenkin}$^{3}$\footnote{email address:
vilenkin@cosmos2.phy.tufts.edu}

\vspace{4mm}
$^1$Theory Division, CERN, CH-1211 Geneva 23, Switzerland \\
$^2$Department of Physics and Astronomy, University of Pennsylvania \\ 
Philadelphia, PA 19104-6396, USA \\
$^3$Institute of Cosmology, Department of Physics and Astronomy \\
Tufts University, Medford, MA 02155, USA

\vspace{20mm}

{\bf Abstract}
\end{center}

A small asymmetry in the flux of neutrinos emitted by a hot newly-born
neutron star could explain the observed motions of pulsars.  However, 
even in the presence of parity-violating processes with anisotropic
scattering amplitudes, no asymmetry is generated in thermal equilibrium. 
We explain why this no-go theorem stymies some of the proposed explanations
for the pulsar ``kick'' velocities.

\vspace{12mm}

\begin{flushleft}
\begin{tabular}{l}
CERN-TH/98-192 \\
May, 1998
\end{tabular}
\end{flushleft}

\vfill

\pagestyle{empty}

\pagebreak

\pagestyle{plain}
\pagenumbering{arabic}
\renewcommand{\thefootnote}{\arabic{footnote}}
\setcounter{footnote}{0}

The observed rapid motions~\cite{cc} of magnetized, rotating neutron stars, 
pulsars, can have a natural explanation if the neutrino emission from a
cooling newly-born neutron star exhibits a small, $\approx$1\% asymmetry.  
One necessary condition for such an asymmetry -- a preferred direction --
is, clearly, satisfied.  A strong magnetic field of the neutron star breaks
the spherical symmetry and can, at least in principle, be responsible for
some anisotropy in neutrino emission.   

One might think that, since the production and scattering of neutrinos in
dense, hot nuclear matter is affected by the magnetic field ${\bf B}$, 
some parity-violating processes could produce an asymmetric flux of
neutrinos in thermal equilibrium.  This conclusion, however, is erroneous. 

In fact, no asymmetry can build up in thermal equilibrium even if the
scattering probabilities are anisotropic.   This was pointed out by one of
the authors in Ref.~\cite{v}.  The argument is similar to that of
Weinberg~\cite{w} with respect to the baryon asymmetry of the Universe.
Our purpose here is to reiterate the discussion of Ref.~\cite{v}
and to elaborate on this general argument in application to
the neutrino emission by a cooling neutron star.   This is particularly
timely because several recent papers~\cite{hl,lq} have reached an erroneous
conclusion with respect to the size of the recoil velocity the pulsars can
receive from neutrino emission.  

The Boltzman equation for neutrinos $\nu$ scattering off neutrons $n$ (it
is straightforward to include the electrons, which we omit for
simplicity) can be written as~\cite{ll}    
\begin{eqnarray}
q_0 \frac{\dd}{\dd t} \fnu ({\bf q},t) & = &  
\sum_{{\bf s},{\bf s}'} \int \frac{d^3 {\bf p}}{p_0} \frac{d^3 {\bf p}'}{p_0'}
\frac{d^3 {\bf q}'}{q_0'} \  
[ \fnu({\bf q}',t)\, \fn_{{\bf s}'}({\bf p}',t) \, W({\bf p}', {\bf
  s}',{\bf q}' | {\bf p}, {\bf s}, {\bf q} )  \nonumber \\ & &  
- \fnu({\bf q},t) \, \fn_{{\bf s}}({\bf p},t) \, W({\bf p}, {\bf s}, {\bf
  q} | {\bf p}', {\bf s}', {\bf q}') ].  
\label{bmn}
\end{eqnarray}
Here $\fnu({\bf q})$ and $\fn({\bf p})$ are the neutrino and neutron
distribution functions, respectively; ${\bf s}$ denotes a neutron spin; and
$W({\bf p}', {\bf s}',{\bf q}' | {\bf p}, {\bf s}, {\bf q} )$ is the
probability of scattering $|n({\bf p},{\bf s}) \nu({\bf q}) \rangle \to |
n({\bf   p}',{\bf s}') \nu({\bf q}') \rangle $ per unit time per unit
phase-space volume. The states are normalized for the invariant phase-space
volume $d^3   {\bf  p}/p_0$ that appears in the integral.    Here we
neglected the effects of fermion degeneracy, the inclusion of which is 
straightforward~\cite{w}.   

We will assume an arbitrary form for the scattering probability $W$, which
may, in particular, be anisotropic.   The essential property of $W$ is
unitarity, 
\beq
\sum_{{\bf s}'} \int \frac{d^3 {\bf p'}}{p'_0} \frac{d^3 {\bf
    q}'}{q_0'} \  W({\bf p}, {\bf s}, {\bf q} | {\bf p}', {\bf s}', {\bf
  q}') = 
\sum_{{\bf s}'} \int \frac{d^3 {\bf p'}}{p'_0} \frac{d^3 {\bf
    q}'}{q_0'} \  W({\bf p}', {\bf s}', {\bf q}' | {\bf p}, {\bf s}, {\bf
  q}) =
1,
\label{uni}
\eeq
which is merely a requirement that the probability is conserved: with
probability one, every initial state scatters into one of the final states, 
and vice versa. 

The unitarity of the scattering matrix, as expressed by
equation~(\ref{uni}), is sufficient to show that the isotropic
time-independent equilibrium distributions, which depend only on energy,
satisfy the Boltzman equation (\ref{bmn}), regardless of any asymmetries in
$W$.  In statistical equilibrium, $\fnu({\bf q}) \propto \exp \{-q_0/T \}$
and $\fn_{\bf s}({\bf p}) \propto \exp \{-E_{\bf s}({\bf p})\}$, 
where $E_{\bf s}({\bf p}) ={\bf p}^2/2m + g_n \mu_{_B} {\bf s} \cdot {\bf
  B}$. From energy conservation, $q_0+E_{\bf s}({\bf p}) = q_0'+E_{\bf
  s'}({\bf  p'})$, so that  
\beq
\fnu(q) \fn_{\bf s}(p) = \fnu(q') \fn_{\bf s'}(p') .
\label{db}
\eeq
Relation (\ref{db}) is quite general and can be regarded as an
expression of detailed balance in statistical equilibrium~\cite{ll}.  

One can now prove that a time-independent spherically-symmetric distribution
function, which satisfies the condition of statistical equilibrium
(\ref{db}), is a solution of the Boltzman equation (\ref{bmn}) for any form 
of $W$. Indeed, if $\fnu $ is independent of time, the left-hand side of
equation  (\ref{bmn}) vanishes.  The right-hand side,  
\beq
\sum_{{\bf s},{\bf s}'} \int \frac{d^3 {\bf p}}{p_0} \frac{d^3 {\bf
    p}'}{p_0'} \frac{d^3 {\bf q}'}{q_0'} \  
\fnu({\bf q})\, \fn_{{\bf s}}({\bf p}) \, [ W({\bf p}', {\bf
  s}',{\bf q}' | {\bf p}, {\bf s}, {\bf q} )  
- \, W({\bf p}, {\bf s}, {\bf
  q} | {\bf p}', {\bf s}', {\bf q}') ], 
\eeq
also vanishes because 
\beq
\sum_{\bf s'} \int \frac{d^3 {\bf
    p}'}{p_0'} \frac{d^3 {\bf q}'}{q_0'} \   [ W({\bf p}', {\bf
  s}',{\bf q}' | {\bf p}, {\bf s}, {\bf q} )  
- \, W({\bf p}, {\bf s}, {\bf
  q} | {\bf p}', {\bf s}', {\bf q}') ] =0 
\eeq
by virtue of equation (\ref{uni}).  So, if the neutrinos are thermally
produced in equilibrium, no asymmetry is generated by an anisotropic
scattering probability. 

It is clear why the analysis of Ref.~\cite{hl} is in contradiction with the
no-go theorem we have proven.  The expression for the scattering
cross-section in equation (3) of Ref.~\cite{hl} explicitly violates unitarity
because it depends asymmetrically on the scattering angle of the outgoing
neutrino and has no dependency on the initial momentum of the incident
neutrino.  In our notation, this corresponds to $W \propto (q_0+ \langle
{\bf s} \rangle \cdot {\bf q}')$, where $\langle {\bf s} \rangle$ is the
average neutron polarization.   
Since this expression does not satisfy the constraints of unitarity, it can
lead to a neutrino asymmetry and hence a pulsar kick. A similar mistake was
made in Ref.~\cite{lq}, where it was also 
claimed that the asymmetry in the distribution of outgoing neutrinos is
proportional to the optical depth of the neutron star, that is to the size
of the region where neutrinos are in thermal equilibrium with nuclear
matter.  It should be clear from our discussion, as well as from the general
principles of thermodynamics, that the size of a system in equilibrium does
not affect the distribution functions. 

Of course, a hot neutron star that emerges from a supernova explosion is
not fully in equilibrium, and this causes some asymmetry in the flux of
outgoing neutrinos~\cite{v}.  The departure from thermal equilibrium is
due to the variation of the macroscopic parameters, such as 
${\bf B}$, temperature, and the matter density, which occur on some length
scale $L$. Therefore, the asymmetry resulting from the non-equilibrium
behavior is proportional to $\lambda/L$, where $\lambda$ is the neutrino
mean free path.  

In addition, an asymmetry in neutrino emission can arise from some other 
interactions of neutrinos at or near the neutrinosphere, as suggested in
Refs.~\cite{ks}.  Since at that point the neutrinos are free-streaming,
they are already out of equilibrium and the above no-go theorem doesn't
apply.   

To summarize, any attempt to explain the pulsar kick velocities by a 
``cumulative'' amplification of a small asymmetry in the course
of numerous collisions~\cite{hl,lq} is doomed to failure.  Multiple 
collisions imply that the thermally-produced neutrinos continue to be in
statistical equilibrium.   Therefore, no asymmetry can build up even if the 
scattering probabilities are anisotropic.

\end{document}